\documentclass[aps,superscriptaddress,twocolumn,amsmath,amssymb]{revtex4}
\usepackage{graphics}
\usepackage{graphicx}
\usepackage{epsfig}
\usepackage{dcolumn}
\usepackage{bm}
\usepackage{hyperref}
\usepackage{natbib}
\setcitestyle{super}
\begin{document}
\title{High-temperature quantum oscillations of the Hall resistance in bulk Bi$_2$Se$_3$}
\author{Marco Busch}
\affiliation{Novel Materials Group, Humboldt-Universit\"at zu Berlin, Newtonstra{\ss}e 15, 12489 Berlin, Germany}
\author{Olivio Chiatti}
\affiliation{Novel Materials Group, Humboldt-Universit\"at zu Berlin, Newtonstra{\ss}e 15, 12489 Berlin, Germany}
\author{Sergio Pezzini}
\affiliation{High Field Magnet Laboratory, Radboud University Nijmegen, P.O. box 9010, 6500 GL Nijmegen, Netherlands}
\author{Steffen Wiedmann}
\affiliation{High Field Magnet Laboratory, Radboud University Nijmegen, P.O. box 9010, 6500 GL Nijmegen, Netherlands}
\author{\mbox{Jaime S\'anchez-Barriga}}
\affiliation{\mbox{Helmholtz-Zentrum-Berlin f\"ur Materialien und Energie, Albert-Einstein-Stra{\ss}e 15, 12489 Berlin, Germany}}
\author{Oliver Rader}
\affiliation{\mbox{Helmholtz-Zentrum-Berlin f\"ur Materialien und Energie, Albert-Einstein-Stra{\ss}e 15, 12489 Berlin, Germany}}
\author{Lada V. Yashina}
\affiliation{Department of Chemistry, Moscow State University, Leninskie Gory 1/3, 119991 Moscow, Russia}
\author{Saskia F. Fischer}
\altaffiliation[Correspondence and requests should be addressed to S.F.F.]{ \mbox{(e-mail:} sfischer@physik.hu-berlin.de).}
\affiliation{Novel Materials Group, Humboldt-Universit\"at zu Berlin, Newtonstra{\ss}e 15, 12489 Berlin, Germany}
\date{\today}
\maketitle

{\bf Helically spin-polarized Dirac fermions (HSDF) in protected topological surface states (TSS) are of high interest as a new state of quantum matter. In three-dimensional (3D) materials with TSS, electronic bulk states often mask the transport properties of HSDF. Recently, the high-field Hall resistance and low-field magnetoresistance indicate that the TSS may coexist with a layered two-dimensional electronic system (2DES). Here, we demonstrate quantum oscillations of the Hall resistance at temperatures up to 50 K in bulk Bi$_2$Se$_3$ with a high electron density $n$ of about $2\!\cdot\!10^{19}$~cm$^{-3}$. From the angular and temperature dependence of the Hall resistance and the Shubnikov-de Haas oscillations we identify 3D and 2D contributions to transport. Angular resolved photoemission spectroscopy proves the existence of TSS. We present a model for Bi$_2$Se$_3$ and suggest that the coexistence of TSS and 2D layered transport stabilizes the quantum oscillations of the Hall resistance.}

Among the new material class of topological insulators (TI), the chalcogenide semiconductor Bi$_2$Se$_3$ has been long subject to intense investigations due to its potential integration in room temperature applications, such as dissipationless electronics and spintronics devices\cite{Xue,Zhang2009,Ando-2013-JPSJ,Chiatti}. Bi$_2$Se$_3$ has a single Dirac cone at the $\Gamma$-point in the first surface Brillouin zone and a direct band gap of 0.3 eV between the valence and the conduction band\cite{Checkelsky,Betancourt,xia09}. Due to the inversion symmetry in Bi$_2$Se$_3$ the topological $Z_2$ invariant $\nu\!=\!(1;000)$ is equal to the charge of parity of the valence band eigenvalues at the time-reversal-invariant points of the first Brillouin zone caused by the band inversion\cite{Fu}. In the crystalline modification Bi$_2$Se$_3$ has a tetradymite structure with R$\overline{3}$m symmetry. The unit cell consists of 15 atomic layers grouped in three quintuple layers (QLs) with Se$-$Bi$-$Se$-$Bi$-$Se order stacked in an A$-$B$-$C$-$A$-$B$-$C manner. The quintuple layers are van der Waals bonded to each other by a double layer of Se atoms, the so-called van der Waals gap\cite{Ando-2013-JPSJ}. The existence of TSS in Bi$_2$Se$_3$ has been experimentally confirmed through angle resolved photoemission spectroscopy (ARPES)\cite{xia09,Bianchi,Chiatti} and scanning tunneling microscopy/scanning tunneling spectroscopy (STM/STS)\cite{Alpichshev,Liu2014}. The as-grown crystals of Bi$_2$Se$_3$ are typically $n$-type because of electron doping due to natural selenium vacancies\cite{Hyde,Felser}. Therefore, the transport properties of Bi$_2$Se$_3$ are generally dominated by bulk conduction. In particular, the temperature dependence of the electrical resistivity $\rho$ is metallic-like\cite{Chiatti,Analytis,Eto,Petrushevsky-2012-prb,caoPRL} and Shubnikov-de Haas (SdH) oscillations in the longitudinal resistivity $\rho_\mathrm{xx}$ show the characteristic signatures for a 3D Fermi surface\cite{Analytis,Eto}. For samples with lower carrier density $n\!\sim\!10^{16}$~cm$^{-3}$, the TSS can be detected via additional SdH oscillations with a frequency $B_\mathrm{SdH}$ higher than that of the bulk\cite{Analytis2010}. The Hall resistivity $\rho_\mathrm{xy}$ exhibits quantum oscillations for a carrier density $n\!<\!5\cdot10^{18}$~cm$^{-3}$. Recently, for $n\!\geq\!2\cdot10^{19}$~cm$^{-3}$ a bulk quantum Hall effect (QHE) with 2D-like transport behavior was reported\cite{Chiatti,Eto,caoPRL}. Its origin remains unidentified.

In this work we demonstrate that the quantum oscillations of the Hall resistance $R_\mathrm{xy}$ in high-purity, nominally undoped Bi$_2$Se$_3$ single crystals with a carrier density of $n\!\approx\!2\cdot10^{19}$~cm$^{-3}$ persists up to high temperatures. The longitudinal resistivity $\rho_\mathrm{xx}$ and the Hall resistivity $\rho_\mathrm{xy}$ were measured simultaneously in a temperature range between 0.3~K and 72~K in tilted magnetic fields up to 33~T. The SdH oscillations in $\rho_\mathrm{xx}$ are clearly dominated by 3D bulk carriers. The quantum oscillations of the Hall resistance $R_\mathrm{xy}$ scales with the sample thickness, strongly indicating 2D layered transport. These findings stand out because the Bi$_2$Se$_3$ sample investigated here has a lower carrier mobility $\mu$ of about $600$~cm$^2$/(Vs) than materials hosting a typical 2D Fermi gas\cite{Beenakker,Stoermer,Novoselov2007,Khouri} or 3D Fermi gas\cite{Halperin,Hannahs,hill-1998-prb} showing QHE. We discuss the conditions of the QHE below in detail and present a model for the coexistence of TSS and 2D layered transport.

\begin{figure*}[t!]
\begin{center}
\includegraphics[width=18cm]{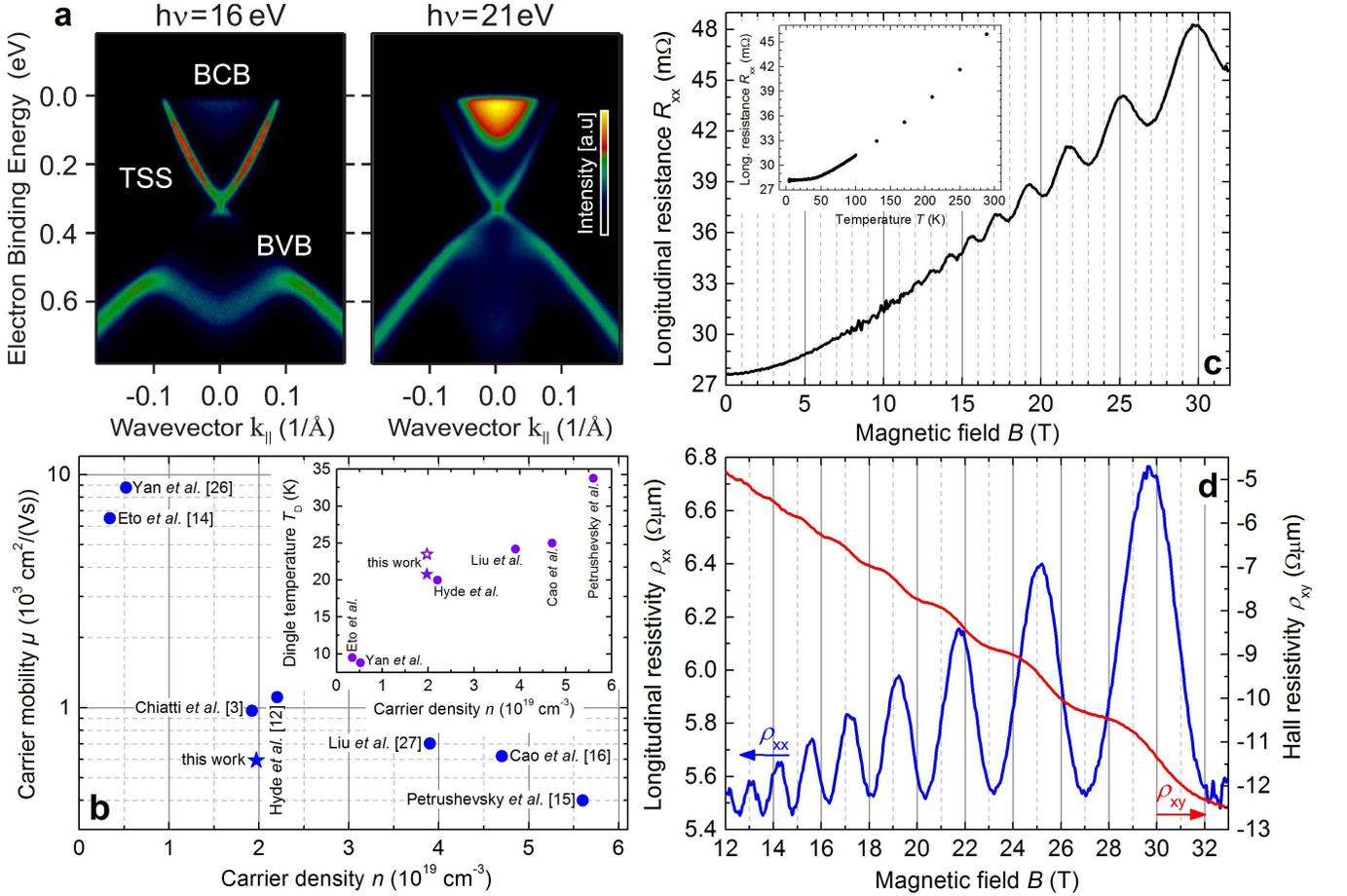}
\end{center}
\vspace{-0.5cm}
\caption{{\bf Electronic structure, temperature-dependent resistance and magnetotransport properties of Bi$_2$Se$_3$. a}, Electronic structure of the Bi$_2$Se$_3$ bulk single crystal before mechanical exfoliation. The panels show high resolution ARPES $E(k_{||})$ dispersions measured at a temperature of $T\!=\!12$~K and at a photon energy of h$\nu\!=\!16$~eV and 21~eV, respectively. In the left panel, the TSS, the bulk conduction band (BCB) and the bulk
valence band (BVB) are indicated. {\bf b}, Carrier mobility $\mu$ vs carrier density $n$ of various crystalline Bi$_2$Se$_3$ samples deduced in different experimental investigations. Inset: Dingle temperature $T_\mathrm{D}$ vs carrier density $n$. The full star indicates the value assuming 2D transport (cf. inset of Fig.~\ref{figure4}c and text) and the open star indicates the value assuming 3D transport (cf. Fig.~\ref{figure4}d and text). {\bf c}, Longitudinal resistance $R_\mathrm{xx}$ vs perpendicular magnetic field $B$ as symmetrized raw data measured at $T\!=\!0.47$~K. Inset: Longitudinal resistance $R_\mathrm{xx}$ vs temperature $T$ measured for $B\!=\!0$. {\bf d}, Longitudinal resistivity $\rho_\mathrm{xx}$ (blue curve, left axis) and Hall resistivity $\rho_\mathrm{xy}$ (red curve, right axis) vs perpendicular magnetic field $B$ measured at $T\!=\!0.47$~K.}
\label{figure1}
\end{figure*}

\begin{figure*}[t!]
\begin{center}
\includegraphics[width=18cm]{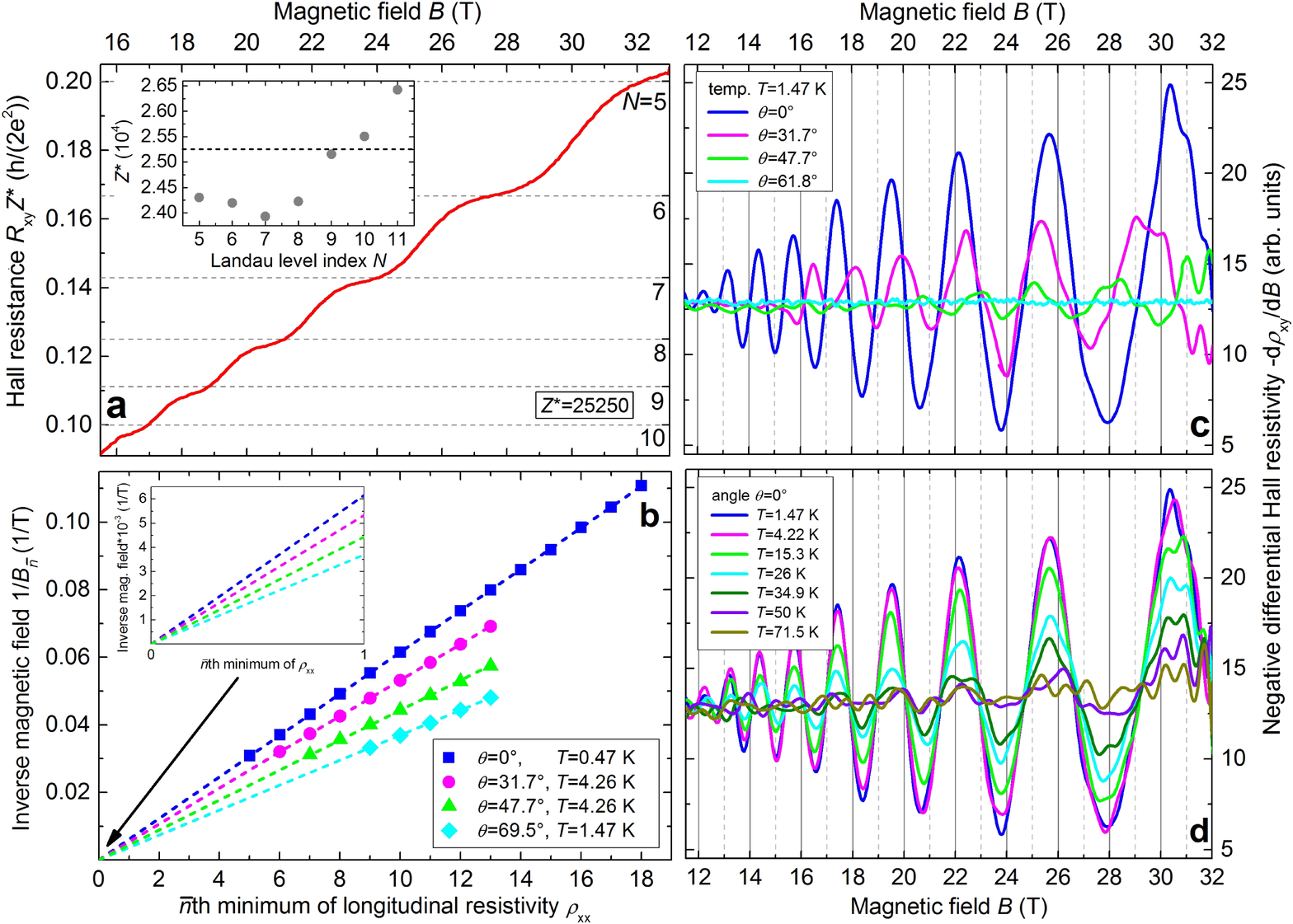}
\end{center}
\vspace{-0.5cm}
\caption{{\bf Quantum oscillations of the Hall resistance. a}, Hall resistance $\widetilde{R}_\mathrm{xy}\!=\!R_\mathrm{xy}Z^*$ in units of $h/\!(2e^2)$ with an averaged number of 2D layers $Z^*\!=\!25250$ vs magnetic field $B$ at $T\!=\!0.47$~K and $\theta\!=\!0^\circ$. Inset: Number of 2D layers $Z^*$ vs Landau level (LL) index $N$. The averaged number of 2D layers $Z^*\!=\!25250$ is shown as dashed line. {\bf b}, LL fan diagram from SdH oscillations of longitudinal resistivity $\rho_\mathrm{xx}$ for different values of the angle $\theta$ between the direction of the magnetic field $\vec{B}$ and the surface normal $\vec{N}$ of the Bi$_2$Se$_3$ macro flake and for different temperatures $T$. The dashed lines represent the best linear fits to the data. Inset: Enlargement of the LL fan diagram for $0\!\leq\!\bar{n}\!\leq\!1$. {\bf c} and {\bf d}, Negative differential Hall resistivity $-\mathrm{d}\rho_\mathrm{xy}/\mathrm{d}B$ vs magnetic field $B$, for different values of the angle $\theta$ at $T\!=\!1.47$~K and at different temperatures $T$ for $\theta\!=\!0^\circ$, respectively.}
\label{figure2}
\end{figure*}

\begin{figure*}[t!]
\begin{center}
\includegraphics[width=18cm]{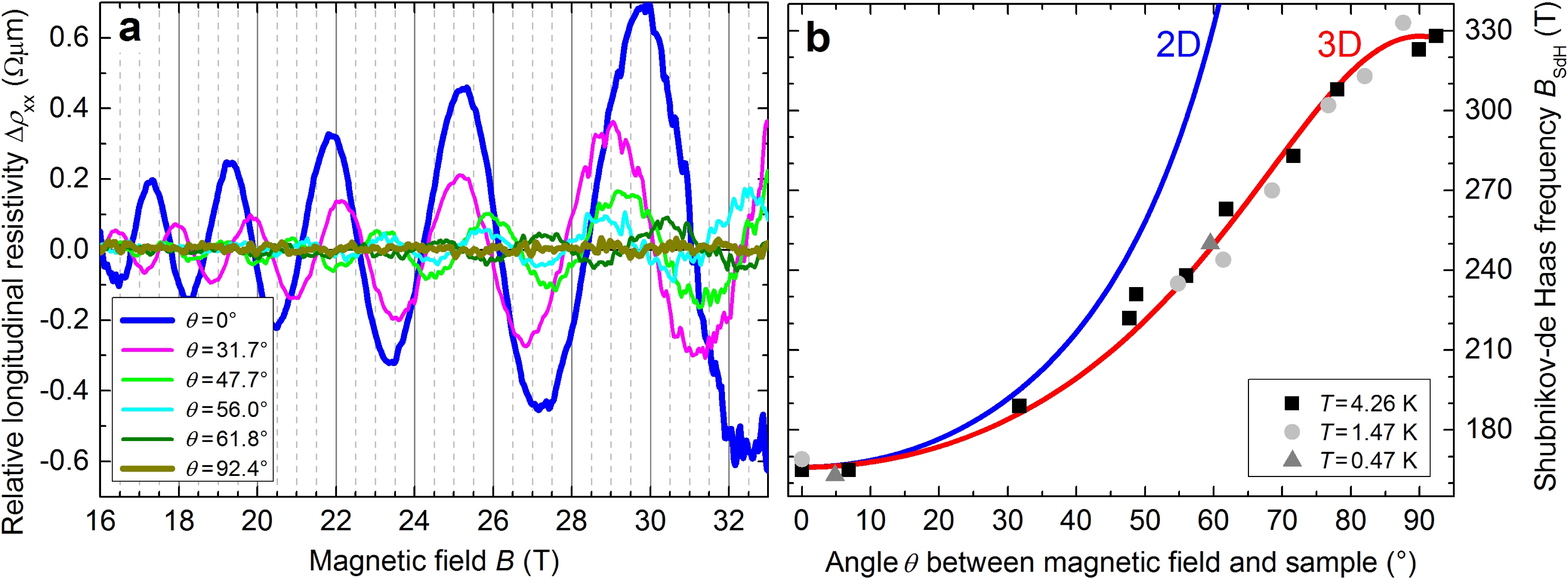}
\end{center}
\vspace{-0.5cm}
\caption{{\bf Angular-dependence of Shubnikov-de Haas oscillations. a}, Relative longitudinal resistivity $\Delta\rho_\mathrm{xx}$ vs magnetic field $B$ measured at $T\!=\!4.26$~K for different values of the angle $\theta$. {\bf b}, SdH frequency $B_\mathrm{SdH}$ vs angle $\theta$ determined at temperatures $T\!=\!0.47$~K (dark gray triangles), 1.5~K (light gray circles), and 4.26~K (black squares). Curves represent calculated behavior for a planar 2D Fermi surface assuming $B_\mathrm{SdH}^\mathrm{2D}\!=\!B_\bot/\cos\theta$ (blue curve) and for an ellipsoidal 3D Fermi surface assuming $B_\mathrm{SdH}^\mathrm{3D}\!=\!B_\bot B_{||}/\sqrt{(B_\bot\cos\theta)^2+(B_{||}\sin\theta)^2}$ (red curve) with $B_\bot\!=\!166$~T (for $\theta\!=\!0^\circ$) and $B_{||}\!=\!328$~T (for $\theta\!=\!90^\circ$).}
\label{figure3}
\end{figure*}

\begin{figure*}[t!]
\begin{center}
\includegraphics[width=18cm]{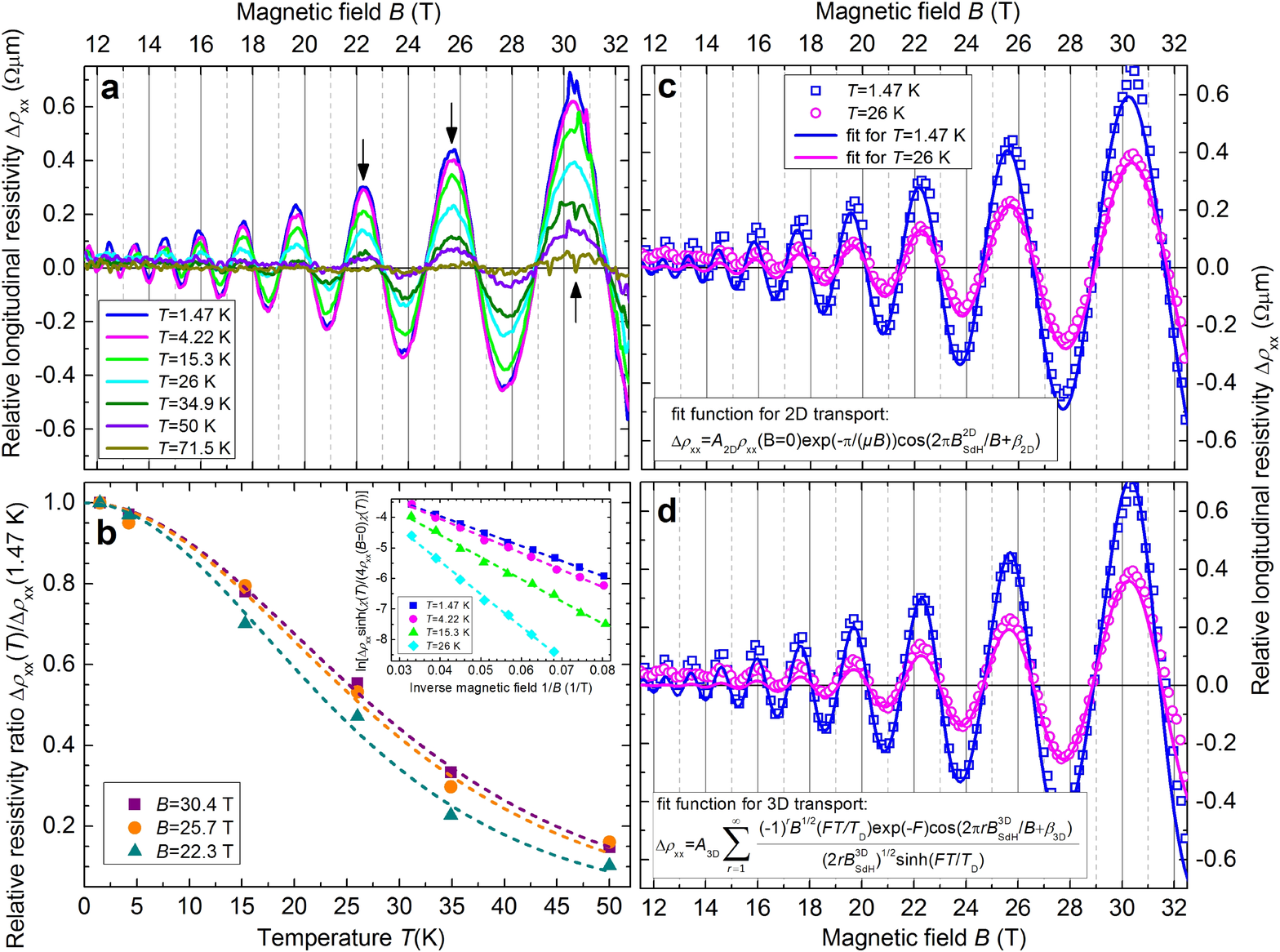}
\end{center}
\vspace{-0.5cm}
\caption{{\bf 2D and 3D analysis of the temperature dependence of the Shubnikov-de Haas oscillations. a}, Relative longitudinal resistivity $\Delta\rho_\mathrm{xx}$ vs magnetic field $B$ measured for an angle $\theta\!=\!0^\circ$ at different temperatures $T$. The black arrows indicate values of magnetic field $B$ shown in panel {\bf b} vs temperature $T$. {\bf b}, Relative longitudinal resistivity ratio $\Delta\rho_\mathrm{xx}(T)/\Delta\rho_\mathrm{xx}(1.47~\mathrm{K})$ vs temperature $T$ for a magnetic field of $B\!=\!30.4$~T (violet squares), 25.7 T (orange circles), and 22.3 T (dark cyan triangles). Dashed curves represent best fits to data assuming the function $\chi(T)/\sinh(\chi(T))$, with $\chi(T)\!=\!(4\pi^3m^*k_\mathrm{B}T)/(heB)$. Inset: Dingle plots of the SdH oscillations (maxima of relative longitudinal resistivity $\Delta\rho_\mathrm{xx}$ as shown in panel (a)) at $T\!=\!1.47$~K (blue squares), 4.22~K (pink circles), 15.3~K (green triangles), and 26~K (cyan diamonds). Dashed lines represent best linear fits to data with the function $-\pi m^*\!/(e\tau_\mathrm{D}B)$, with $m^*\!=\!0.16m_\mathrm{e}$. {\bf c} and {\bf d}, Relative longitudinal resistivity $\Delta\rho_\mathrm{xx}$ vs magnetic field in the range 12 T$\,\leq\!B\!\leq\!33$~T measured for an angle $\theta\!=\!0^\circ$ at $T\!=\!1.47$~K (blue squares) and 26~K (pink circles). For clarity, only every fifth data point is shown. Panel {\bf c}: Curves represent best fits assuming 2D transport, using the fit function given in the legend with the SdH frequency $B_\mathrm{SdH}^\mathrm{2D}\!=\!166$~T (cf. Fig.~\ref{figure3}b and text). Panel {\bf d}: Curves represent best fits assuming 3D transport, using the fit function given in the legend with the SdH frequency $B_\mathrm{SdH}^\mathrm{3D}\!=\!169.5$~T and the parameter $F\!=\!2\pi rk_\mathrm{B}T_\mathrm{D}/(\hbar\omega_\mathrm{C})\!=\!rm^*/(\tau_\mathrm{D}eB)$ with $m^*\!=\!0.16m_\mathrm{e}$. The parameter $r$ denotes the number of harmonic oscillations. In the present study we considered a range of values of $1\!\leq\!r\!\leq\!20$.}
\label{figure4}
\end{figure*}

~\\{\large{\bf$\!\!\!$Results}}\\
{\bf Experimental data.} In Fig. 1a, we show high-reso-\ lution ARPES dispersions measured at a temperature of 12~K for two representative photon energies of h$\nu\!=\!16$~eV and 21~eV. We clearly observe distinct intensity contributions from the bulk conduction band (BCB) and bulk valence band (BVB) coexisting with sharp and intense Dirac cone representing the TSS. The BCB crossing the Fermi level indicates that the crystals are intrinsically $n$-type, in agreement with our Hall measurements on the same samples. At binding energies higher than the Dirac node ($E_\mathrm{D}\!\sim\!0.35$~eV), the lower half of the TSS overlaps with the BVB. By changing the photon energy we select the component of the electron wave vector perpendicular to the surface $k_z$. Since the lattice constant of Bi$_2$Se$_3$ is very large along the $z$ direction ($c\!=\!28.64$~\AA), the size of the bulk Brillouin zone (BBZ) is very small ($\sim\!0.2$~\AA$^{-1}$). With photon energies between 16 to 21~eV we cross practically the complete BBZ, enhancing the sensitivity to the out-of-plane dispersion of the bulk bands. We note that the ARPES intensity changes with the photon energy as well due to the $k_z$-dependence of the photoemission transitions. Differently from the BCB or BVB, the TSS exhibits no $k_z$-dependence due to its 2D character. Consistent with the direct nature of the gap, we find the BCB minimum at 21~eV ($\approx\!\Gamma$-point of the BBZ) and at a binding energy of $\sim\!0.154$ eV, while the BVB maximum at $\sim\!0.452$ eV. In particular, from the ARPES measurements, we
estimate a bulk carrier density of $n_\mathrm{3D,BCB}\!=\!1.77\cdot10^{19}$~cm$^{-3}$ and a sheet carrier density of $n_\mathrm{2D,TSS}\!=\!k_{\rm F,TSS}^2/(4\pi)\!=\!1.18\cdot10^{13}$~cm$^{-2}$, with $k_\mathrm{F,3D}\!=\!0.064$~\AA$^{-1}$ and $k_{\rm F,TSS}\!=\!(0.086\pm0.001)$~\AA$^{-1}$, respectively.

The temperature-dependent longitudinal resistance $R_\mathrm{xx}$ at zero magnetic field shows metallic-like behavior (see inset of Fig.~\ref{figure1}c). A residual resistance ratio $\mathrm{RRR}\!=\!R_{\mathrm{xx}}(288~\mathrm{K})/R_{\mathrm{xx}}(4.3~\mathrm{K})\!=\!1.63$ indicates a high crystalline quality\cite{Hyde} (see Supplementary Information Sec. 1). The resistance $R_{\mathrm{xx}}$ remains practically constant from very low temperatures up to 30 K, presumably due to a combination of surface states and static disorder scattering of the charge carriers. These observations are in agreement with previous reports\cite{Chiatti,Eto,caoPRL,Petrushevsky-2012-prb,Yan2014,Liu2015}. 

SdH oscillations in the longitudinal resistivity $\rho_\mathrm{xx}$ and the onset of quantum oscillations of the Hall resistivity $\rho_\mathrm{xy}$ can be observed at magnetic fields $B\!\geq\!10$~T. Fig.~\ref{figure1}d depicts $\rho_\mathrm{xx}$ (blue curve, left axis) and $\rho_\mathrm{xy}$ (red curve, right axis) as a function of the perpendicular magnetic field $B$ at a temperature of $T\!=\!0.47$~K. In Fig.~\ref{figure1}c we show the corresponding longitudinal resistance $R_\mathrm{xx}$ vs perpendicular magnetic field $B$ measured at $T\!=\!0.47$~K as symmetrized raw data $R_\mathrm{xx}^\mathrm{sym}(B)\!=\!\left[R_\mathrm{xx}^\mathrm{raw}(+B)+R_\mathrm{xx}^\mathrm{raw}(-B)\right]/2$. The slope of the Hall resistivity $\rho_\mathrm{xy}$ yields a carrier mobility of $\mu_\mathrm{Hall}\!=\!594$ cm$^2$/(Vs) and a carrier density of $n_\mathrm{Hall}\!=\!1.97\cdot10^{19}$~cm$^{-3}$. In Fig.~\ref{figure1}b we compare the carrier mobility $\mu$ vs carrier density $n$ of various crystalline Bi$_2$Se$_3$ samples, obtained in magnetotransport measurements. Our data are in the medium carrier density and mobility range.

Fig.~\ref{figure2}a details the analysis performed on the high-field anti-symmetrized $R^\mathrm{asy}_\mathrm{xy}(B)\!=\!\left[R_\mathrm{xy}^\mathrm{raw}(+B)-R_\mathrm{xy}^\mathrm{raw}(-B)\right]/2$ data for $T\!=\!0.47$~K and an angle of $\theta\!=\!0^\circ$. $\theta$ denotes the angle between the direction of $\vec{B}$ and the surface normal $\vec{N}$ and the $c$-axis of the Bi$_2$Se$_3$ macro flake (i.e. $\theta\!=\!0^\circ$ means $\vec{B}\,||\,\vec{N}$). The scaling behaviour of $\Delta R^\mathrm{asy}_\mathrm{xy}\!=\!R^\mathrm{asy}_\mathrm{xy}(N)-R^\mathrm{asy}_\mathrm{xy}(N+1)$ with the thickness leads to $Z^*\!=\!\left[(1/N-1/(N+1))/\Delta R^\mathrm{asy}_\mathrm{xy}\right]\cdot(h/(2e^2))$ as the number of 2D spin-degenerate layers contributing to the transport. Conclusively, an average number of 2D layers of $Z^*\!=\!25250$ is derived. The variation of $Z^*$ for different Landau level (LL) index $N$ is given in inset of Fig.~\ref{figure2}a.

Figs.~\ref{figure2}c and \ref{figure2}d depict the angular and the temperature dependence, respectively, of the negative differentiated Hall resistivity $-\mathrm{d}\rho_\mathrm{xy}/\mathrm{d}B$ vs magnetic field $B$, determined for different angles $\theta$ at $T\!=\!1.47$~K, as well as different temperatures $T$ for $\theta\!=\!0^\circ$. In accordance with the angular and the temperature dependence of the SdH oscillations as shown in Figs.~\ref{figure3}a and \ref{figure4}a, respectively, a decreasing amplitude of the differentiated Hall resistivity with increasing angle $\theta$ and increasing temperature $T$ is detected. At a constant temperature of $T\!=\!1.47$~K the typical signatures of quantum oscillations of the Hall resistance are observed up to an angle of $61.8^\circ$ and at $\theta\!=\!0^\circ$ the amplitude of $\mathrm{d}\rho_\mathrm{xy}/\mathrm{d}B$ vanishes only for temperatures above 71.5~K.

In the LL fan diagram shown in Fig.~\ref{figure2}b the inverse magnetic-field positions $1/B_{\bar{n}}$ are plotted vs the $\bar{n}$th minimum of the longitudinal resistivity $\rho_\mathrm{xx}$ for different values of $\theta$. The straight dashed lines, which represent the best linear fits to the data, intersect jointly the $\bar{n}$-axis at the point of origin (see Supplementary Information Sec. 2). Hence, we find for all angles $\theta$ and temperatures $T$ investigated here a significant evidence for a trivial Berry phase of $\Phi_\mathrm{B}\!=\!0$ (cf. inset of Fig.~\ref{figure2}b) and conclude the dominance of non-relativistic fermions. For an improved estimation of the trivial Berry phase $\Phi_\mathrm{B}\!=\!0$ we have fitted the behavior of the relative longitudinal resistivity $\Delta\rho_\mathrm{xx}$ vs magnetic field $B$ assuming 2D and 3D transport (cf. Fig.~\ref{figure4}c and d, respectively).

In Fig.~\ref{figure3}a we show the relative longitudinal resistivity $\Delta\rho_\mathrm{xx}$ vs magnetic field $B$ measured at $T\!=\!4.26$~K for different angles $\theta$. The relative longitudinal resistivity $\Delta\rho_\mathrm{xx}$ was calculated from the measured longitudinal resistivity $\rho_\mathrm{xx}$ by subtracting a suitable polynominal fit to the background, in order to extract the oscillatory component. The amplitude of the SdH oscillations decreases with increasing angle $\theta$, and is really marginal for $\theta\!>\!70^\circ$. Furthermore, a shift of the SdH oscillations with increasing angle $\theta$ with respect to $\theta\!=\!0^\circ$ is observed. For all values of $\theta$ and $T$, we found one value of the SdH frequency $B_\mathrm{SdH}$. The absence of additional frequencies and beatings, as well as the angular dependence of $B_\mathrm{SdH}$ (see Fig.~\ref{figure3}b), are significant evidence of a single 3D (non-spherical) Fermi surface (see Supplementary Information Sec. 3). 

The temperature dependence of $\Delta\rho_\mathrm{xx}$ is shown in Fig.~\ref{figure4}a: the amplitude decreases with increasing temperature $T$, and oscillations are not observed for $T\!>\!71.5$~K. From the fitting of the relative longitudinal resistivity ratio $\Delta\rho_\mathrm{xx}(T)/\Delta\rho_\mathrm{xx}(T\!=1.47$ K), we deduce an effective mass of the charge carriers of $m^*\!\cong\!0.16m_\mathrm{e}$ ($m_\mathrm{e}\!=\!9.10938356\cdot10^{-31}$~kg denotes the electron rest mass) and a Fermi velocity of $v_\mathrm{F}\!=\!\hbar k_\mathrm{F,3D}/m^*\!=\!0.46\cdot10^6$~m/s, with $k_\mathrm{F,3D}\!=\!0.064$~\AA$^{-1}$.

The Dingle plots (inset of Fig.~\ref{figure4}b) at temperatures of $T\!=\!1.47$~K, 4.22~K, 15.3~K, and 26~K yield the following Dingle scattering time (also known as single-particle relaxation time) $\tau_\mathrm{D}$ and the Dingle temperature $T_\mathrm{D}\!=\!h/(4\pi^2 k_\mathrm{B}\tau_\mathrm{D})$, assuming the fit function $-\pi m^*/(e\tau_\mathrm{D}B)$ with $m^*\!=\!0.16m_\mathrm{e}$: $\tau_\mathrm{D}\!=\!5.8\cdot10^{-14}$~s ($T_\mathrm{D}\!=\!20.8$~K), $5.1\cdot10^{-14}$~s (23.7~K), $3.9\cdot10^{-14}$~s (30.9~K) and $2.7\cdot10^{-14}$~s (45.5~K), respectively.

For a more detailed analysis we have fitted the magnetic-field dependence of $\Delta\rho_\mathrm{xx}$ (see Supplementary Information Sec. 4). In a first step we assumed 2D transport in accordance with other investigations\cite{Eto,Petrushevsky-2012-prb,Yan2014} and have used as fit function the Lifshitz-Kosevich formula\cite{Ando-2013-JPSJ,Lifshitz,Taskin}. Although we found a reasonably good agreement between experimental data and the calculated behavior for $\Delta\rho_\mathrm{xx}(B)$ under the assumption of 2D transport (cf. Fig.~\ref{figure4}c), we also performed fits under the assumption of 3D transport (cf. Fig.~\ref{figure4}d). This is motivated by the analysis of the angular dependence of the SdH frequency $B_\mathrm{SdH}$ shown in Fig.~\ref{figure3}b, were we found an ellipsoidal 3D Fermi surface. We find for all curves a single value for the Dingle temperature $T_\mathrm{D}\!=\!23.5$~K and hence a single value for the Dingle scattering time $\tau_\mathrm{D}\!=\!5.2\cdot10^{-14}$~s, consistent with a constant $R_\mathrm{xx}(T)$ up to $T\!=\!30$~K (see inset of Fig.~\ref{figure1}c). From $\tau_\mathrm{D}$ and the effective mass $m^*\!=\!0.16m_\mathrm{e}$, we determined a carrier mobility of $\mu_\mathrm{D}\!=\!e\tau_\mathrm{D}/m^*\!=\!572$~cm$^2$/(Vs). 

{\bf Evaluation of experimental data.} The angle dependence of the SdH oscillations was used to determine the ellipsoidal shape of the Fermi surface. For a plane 2D Fermi surface, the SdH oscillation frequency is equal to $B_\mathrm{SdH}^\mathrm{2D}\!(\theta)=\!B_\bot/\cos\theta$, with $B_\mathrm{SdH}^\mathrm{2D}(\theta)\!\rightarrow\!\infty$ for $\theta\!\rightarrow\!90^\circ$ (blue curve in Fig.~\ref{figure3}b) and for an ellipsoidal 3D Fermi surface it is $B_\mathrm{SdH}^\mathrm{3D}(\theta))\!=\!B_\bot B_{||}/\sqrt{(B_{||}\cos\theta)^2+(B_\bot\sin\theta)^2}$ (red curve in Fig.~\ref{figure3}b), with $B_\bot\!=\!B_\mathrm{SdH}^\mathrm{3D}(\theta\!=\!0^\circ)\!=\!B_\mathrm{SdH}^\mathrm{2D}(\theta\!=\!0^\circ)\!=\!166$~T and $B_{||}\!=\!B_\mathrm{SdH}^\mathrm{3D}(\theta\!=\!90^\circ)\!=\!328$~T. Previous data\cite{Petrushevsky-2012-prb,caoPRL} may also be interpreted as 3D ellipsoidal Fermi surface (see Supplementary Information Sec. 3). 

We estimate the ellipsoidal cross-section of the 3D Fermi surface with the wave vectors $k_\mathrm{F,SdH}^\mathrm{(a)}\!=\!k_\mathrm{F,SdH}^\mathrm{(b)}\!=\!\sqrt{2eB_\bot/\hbar}\!=\!0.071$~\AA$^{-1}$ and $k_\mathrm{F,SdH}^\mathrm{(c)}\!=\!2eB_{||}/(\hbar k_\mathrm{F,SdH}^\mathrm{(a)})\!=\!0.14$~\AA$^{-1}$. With these values we deduced an eccentricity for the 3D non-spherical Fermi surface of $k_\mathrm{F,SdH}^\mathrm{(c)}/k_\mathrm{F,SdH}^\mathrm{(a)}\!=\!1.98$. K\"ohler\cite{Koehler} and Hyde {\it et al.}\cite{Hyde} show, that the eccentricity of the non-spherical Fermi surface decreases with decreasing carrier density $n$. In accordance with the present study, Eto {\it et al.}\cite{Eto} deduced for a Bi$_2$Se$_3$ bulk single crystal with a lower carrier density of $n\!=\!3.4\cdot10^{18}$~cm$^{-3}$ an eccentricity of $k^\mathrm{(c)}_\mathrm{F,SdH}/k^\mathrm{(a)}_\mathrm{F,SdH}\!=\!1.62$, consistent with eccentricities obtained by K\"ohler\cite{Koehler}. Assuming a parabolic dispersion and using the values of $k_\mathrm{F}^\mathrm{(a)}$ and $k_\mathrm{F}^\mathrm{(c)}$ from the SdH analysis and of $E_\mathrm{F}$ from the ARPES measurements, we estimate with $E_\mathrm{F}\!=\!(\hbar k_\mathrm{F})^2/(2m^*)$ for the effective masses $m^*_\mathrm{a}\!=\!m^*_\mathrm{b}\!=\!0.125m_\mathrm{e}$ and $m^*_\mathrm{c}\!=\!0.485m_\mathrm{e}$. An average value for the effective mass is then given by\cite{Ziman} $1/m^*\!=\!(1/m_\mathrm{c}^*+2/m_\mathrm{a}^*)/3$, which yields $m^*\!=\!0.166m_\mathrm{e}$. This value is consistent with the value obtained from the temperature dependence of the SdH oscillations: $m_\mathrm{SdH}^*\!=\!0.16m_\mathrm{e}$.   

~\\{\large{\bf$\!\!\!$Discussion}}\\
Generally, a bulk or 3D QHE is attributed to parallel 2D conduction channels, each made from one or a few stacking layers. A bulk QHE, where quantized values of the Hall resistance $R_\mathrm{xy}$ inversely scale with the sample thickness, has been observed in a number of anisotropic, layered electronic bulk materials, e.g., GaAs/AlGaAs multi-quantum wells\cite{Stoermer}, Bechgaard salts\cite{Hannahs,Balicas} and also in Fe-doped Bi$_2$Se$_3$ bulk samples\cite{Ge-2015-ssc}, where transport by TSS was excluded. However, the observation of the quantum oscillations of the Hall resistance in Bi$_2$Se$_3$ at elevated temperatures calls for a special condition considering the usual requirement of $\mu B\!\gg\!1$. In the present case $B_\mathrm{max}\!=\!33$~T and the carrier mobility $\mu\!\approx\!600$~cm$^2$/(Vs) yields only $\mu B_\mathrm{max}\!\approx\!2$. Furthermore, the deduced effective mass $m^*\!=\!0.16m_\mathrm{e}$ yields for a magnetic field of $B\!=\!10$~T, where we observe the onset of the quantum oscillations of the Hall resistance $R_\mathrm{xy}$, a value for the LL energy splitting of $\hbar\omega_\mathrm{c}\!=\!\hbar eB/m^*\!\approx\!7$~meV. However, the thermal energy amounts to $k_\mathrm{B}T\!\approx\!4$~meV at $T\!=\!50$~K, while $\hbar\omega_\mathrm{c}\!\gg\!k_\mathrm{B}T$ is usually required for a QHE. Nevertheless, we observe unambiguous signatures of quantum oscillations in $-\mathrm{d}\rho_\mathrm{xy}/\mathrm{d}B$, as shown in Figs.~\ref{figure2}c and \ref{figure2}d. 

In order to explain the experimental observations, we propose the following model. The Bi$_2$Se$_3$ bulk sample investigated here may consist of three different conducting regions: a semiconducting-like core region, surrounded by a metallic-like shell region and the topological surface (see Fig.~S1 in Supplementary Information Sec. 5). The semiconducting-like core was proven by the preparation of semiconducting micro flakes\cite{Chiatti}. The metallic-like shell region due to Se depletion dominates the transport mechanism observed here as metallic and 2D layered effects. From our experiments, we assume the shell to form a stacked system of 2D layers with a periodic potential\cite{Halperin} either due to the van der Waals-gaps or the unit cell along the $c$-axis because of the carrier density modulation due to Se vacancies. In magnetic fields $B\!\neq\!0$ the thickness scaling of the plateaux-like features in the Hall resistance yields an effective thickness for the shell of stacked 2D layers. For the charge carrier density, we estimate three different values for the core\cite{Chiatti}, the shell and the topological surface: $n_\mathrm{core}\!\approx\!1.2\cdot10^{17}$~cm$^{-3}$, $n_\mathrm{shell}\!\approx\!2\cdot10^{19}$~cm$^{-3}$ and $n_\mathrm{TSS}\!=\!1.2\cdot10^{13}$~cm$^{-2}$, respectively. In the semiconducting-like core region, the Fermi level (chemical potential) is in the gap close to the bottom of the conduction band, whereas in the metallic-like shell region the Fermi level is in the conduction band (see Supplementary Information Sec. 5). 
 
For the SdH frequency $B_\mathrm{SdH}$ we estimate at $\theta\!=\!0^\circ$ for the three regions the following values: $B_\mathrm{SdH,core}\!=\!4.82$~T, $B_\mathrm{SdH,shell}\!=\!166$~T and $B_\mathrm{SdH,TSS}\!=\!248$~T. The small value $B_\mathrm{SdH,core}$ corresponds to a slow-changing background which is out of the measurement range of our experimental setup. The larger value of the TSS is caused only by the small number of surface electrons with respect to the large number of bulk electrons ($N_\mathrm{bulk}\!\approx\!N_\mathrm{shell}\!\approx\!2\cdot10^{15}$ and $N_\mathrm{TSS}\!\approx\!3\cdot10^{11}$ yield a ratio $N_\mathrm{TSS}/N_\mathrm{bulk}\!\approx\!10^{-4}$). Therefore, from the experimental data we deduce only the $B_\mathrm{SdH}$ value for the shell (see Fig.~\ref{figure3}a) and find the dominant contribution of the bulk (core+shell) in the transport behavior. A periodic modulation of the charge carrier density along the $c$-direction would result in a miniband structure for the LLs and, as long as the Fermi level is in a gap between these minibands, the Hall resistivity $\rho_\mathrm{xy}$ will be quantized and scale with the periodicity of the potential\cite{Halperin}. The persistence of the quantum oscillations of the Hall resistance $R_\mathrm{xy}$ up to high temperatures requires the Fermi level pinning in the miniband gap. Therefore, we conclude that the observation of the quantum oscillations of the Hall resistance at higher temperatures in Bi$_2$Se$_3$ with a majority of non-Dirac Fermions is related to the existence of the TSS. Based on our results, we propose that other 3D materials with TSS and a periodic potential modulation may show quantization effects in the Hall resistance at elevated temperatures. 

~\\{\large{\bf$\!\!\!\!$Methods}}\\
High-quality single crystalline Bi$_2$Se$_3$ was prepared from melt with the Bridgman technique. The growth time, including cooling was about 2 weeks for a $\sim$50~g crystal. The whole crystal was easily cleaved along the [00.1] growth direction, indicating crystal perfection. The macro flake was prepared by cleaving the bulk single crystal with a thickness of around 110~$\mu$m to investigate bulk properties.

We explored the structural properties of the bulk single crystal\cite{Chiatti} with atomic force microscopy (AFM), scanning transmission electron microscopy (STEM) and high-resolution transmission electron microscopy (HRTEM). The composition and surface stability were investigated using energy-dispersive x-ray spectroscopy (EDX) and spatially resolved core-level X-ray PEEM. Structural analysis using HRTEM and STEM was carried out at a JEOL JEM2200FS microscope operated at 200~kV. The sample preparation for HRTEM characterization consisted of ultrasonic separation of the flakes from the substrate, followed by their transfer onto a carbon-coated copper grid. Using adhesive tape, the surface was prepared by cleavage of the crystal along its trigonal axis in the direction perpendicular to the van-der-Waals-type $(0001)$ planes. The ARPES measurements were performed at a temperature of 12~K in an ultra-high vacuum (UHV) chamber at a pressure of $\sim5\cdot10^{-10}$ mbar with a VG Scienta R8000 electron analyzer at the UE112-PGM2a beamline of BESSY II using p-polarized undulator radiation. 

We have performed magnetotransport experiments using a standard low-noise Lock-In techniques (Stanford Research Systems SR830 with a Keithley 6221 as current source), with low excitation to prevent heating of the sample. The Bi$_2$Se$_3$ macro flake has been measured in a flow cryostat (1.3~K to 300~K), as well as in a $^3$He insert (down to 0.3~K), in a Bitter magnet with a bore diameter of 32~mm and with magnetic fields up to 33~T at the High Field Magnet Laboratory (HFML) of the Radboud University Nijmegen. In both setups, we used a Cernox thermometer in the vicinity of the sample to monitor the temperature in situ. In the $^3$He system, the temperature between 0.3~K and 1.3~K was stabilized by the $^3$He vapour pressure prior to the magnetic field sweep to assure a constant temperature. However, the temperature between 1.3~K and 4.2~K was stabilized by the $^4$He pressure. Above a temperature of 4.2~K, we have used the flow cryostat and stabilized the temperature using a capacitance.

\section*{Acknowledgements}
Financial support from the Deutsche Forschungsgemeinschaft within the priority program SPP1666 (Grant No. FI932/7-1 and RA1041/7-1) and the Bundesministerium f\"ur Bildung und Forschung (Grant No. 05K10WMA) is gratefully acknowledged.

\section*{Author contributions}
M.B., O.C., S.P., S.W. and S.F.F. contributed to the transport experiments, analyzed the data and wrote the manuscript, J.S.-B. and O.R. conducted the ARPES experiments and L.V.Y. conducted the bulk crystal growth. All authors contributed to the discussion and reviewed the manuscript.

\section*{Additional information}
Supplementary Information is available in the online version of the paper. Reprints and permissions information is available online at www.nature.com/reprints. Correspondence and requests for materials should be addressed to S.F.F.

\section*{Competing financial interests} 
The authors declare no competing financial interests.

\end{document}